\documentclass[twocolumn,pra,aps]{revtex4}
\usepackage{graphicx, color, fancybox} 
\usepackage{amsfonts}
\usepackage{amssymb,amsmath,amsthm}

\begin{document}
%%%%%%%%%%%%%%%%%%%%%%%%%%%%%%%%%%%%%%%%%%%%%%%%%%%%%%%%%%%%%%%%%%%%%%%%%%%%%%
%%% New command %%%
%%%%%%%%%%%%%%%%%%%%%%%%%%%%%%%%%%%%%%%%%%%%%%%%%%%%%%%%%%%%%%%%%%%%%%%%%%%%%%
\newcommand{\ket}[1]{|\,{#1}\,\rangle}
\newcommand{\bra}[1]{\langle\,{#1}\,|}
\newcommand{\braket}[2]{\langle\,#1\,|\,#2\,\rangle}
\newcommand{\mbold}[1]{\mbox{\boldmath $#1$}}
\newcommand{\sbold}[1]{\mbox{\boldmath ${\scriptstyle #1}$}}
\newcommand{\tr}{\,{\rm tr}\,}
%%%%%%%%%%%%%%%%%%%%%%%%%%%%%%%%%%%%%%%%%%%%%%%%%%%%%%%%%%%%%%%%%%%%%%%%%%%%%%
%%% Title %%%
%%%%%%%%%%%%%%%%%%%%%%%%%%%%%%%%%%%%%%%%%%%%%%%%%%%%%%%%%%%%%%%%%%%%%%%%%%%%%%
\title{Discrimination with an error margin among three symmetric states of a qubit}
\author{H.~Sugimoto, Y.~Taninaka, and A.~Hayashi}
\affiliation{Department of Applied Physics, 
           University of Fukui, Fukui 910-8507, Japan}
%%%%%%%%%%%%%%%%%%%%%%%%%%%%%%%%%%%%%%%%%%%%%%%%%%%%%%%%%%%%%%%%%%%%%%%%%%%%%%
%%% Abstract %%%
%%%%%%%%%%%%%%%%%%%%%%%%%%%%%%%%%%%%%%%%%%%%%%%%%%%%%%%%%%%%%%%%%%%%%%%%%%%%%%
\begin{abstract}
We consider a state discrimination problem which deals with settings of minimum-error and unambiguous discrimination systematically by introducing a margin for the probability of an incorrect guess. 
We analyze discrimination of three symmetric pure states of a qubit. 
The  measurements are classified into three types, and one of the three types is optimal depending on the value of the error margin. 
The problem is formulated as one of semidefinite programming. 
Starting with the dual problem derived from the primal one, we analytically obtain the optimal success probability and the optimal measurement that attains it in each domain of the error margin. 
Moreover, we analyze the case of three symmetric mixed states of a qubit. 
\end{abstract}

\maketitle
%%%%%%%%%%%%%%%%%%%%%%%%%%%%%%%%%%%%%%%%%%%%%%%%%%%%%%%%%%%%%%%%%%%%%%%%%%%%%%
%%% Introduction %%%
%%%%%%%%%%%%%%%%%%%%%%%%%%%%%%%%%%%%%%%%%%%%%%%%%%%%%%%%%%%%%%%%%%%%%%%%%%%%%%
\section{Introduction}

In quantum mechanics, it is well known that there is no way to distinguish different nonorthogonal quantum states perfectly without a wrong guess by measurement. 
This is because quantum measurement is statistical in nature and it generally destroys the state of the system to be measured.

Quantum state discrimination \cite{Chefles00}, as with many ideas in quantum information theory, is most easily understood using the metaphor of a game involving two parties, Alice and Bob. 
Alice chooses a state $\rho_{i}$ $(i = 1, 2, \cdots, n)$ from a set of quantum states $\{ \rho_{i} \}$ with some occurrence probabilities known to both parties. 
She gives state $\rho_{i}$ to Bob, whose task is to identify the given state $\rho_{i}$ with one in the set $\{ \rho_{i} \}$. 
When considering such a discrimination problem, two settings are often studied as a standard one. 
In one setting, the discrimination success probability is maximized without any restriction on the probability of an incorrect guess \cite{Helstrom76}. 
This is called minimum-error discrimination since the mean probability of error is minimized as a consequence. 
On the other hand, a wrong guess is not allowed in the setting of unambiguous discrimination. 
Instead, the inconclusive result "I don't know" is permitted when the measurement fails to give a 
definite identification for the input state \cite{Ivanovic87, Dieks88, Peres88, Jaeger95}. 
Some other alternative approaches have also been proposed. 
One interesting scheme is a maximum-confidence measurement analyzed in Refs. \cite{Croke06, Jimenez11, Herzog12}. 
In the other scheme, considered in Refs. \cite{Chefles_J_Mod_Opt98, Zhang99, Fiurasek03, Eldar03}, the probability of correct discrimination is maximized while the rate of inconclusive results is fixed.

We consider a setting of maximizing the discrimination success probability under the condition that the mean probability of error should not exceed a certain error margin $m$ \cite{Touzel07, Hayashi08, Sugimoto09}. 
When the error margin $m$ is 0, the setting is equivalent to unambiguous discrimination. 
In minimum-error discrimination, no condition is imposed on the probability of error, which means the error margin $m$ is 1. 
Thus, this formulation naturally unifies the two commonly adopted settings by controlling the error margin. 
In our previous paper \cite{Sugimoto09}, we analyzed discrimination with the error margin between two pure states with general occurrence probabilities and obtained the optimal discrimination success probability in a fully analytic form. 
The two-dimensional parameter space consisting of occurrence probabilities and the error margin is divided into three domains: minimum-error, intermediate, and single-state domain. 
The types of optimal measurement differ depending on the domain. 
However, for the discrimination problem among more than two pure states, even in the two standard settings it is not easy to obtain analytical solutions, though a great number of works on general theories have been reported \cite{Andersson02, Hunter04, Barnett09, Samsonov2009, Assalini10, Chefles98, Chefles_Barnett98, Peres98, Sun01, Zhang01, Eldar_IEEE03, Jafarizadeh08, Samsonov09, Pang09, Sugimoto10}. 

In this paper, we consider the case of three symmetric states \cite{Chefles98} of a qubit. 
We formulate the discrimination problem among three quantum states with general occurrence probabilities in Sec. \ref{sec:formulation}. 
Then we analyze the case of three symmetric states of a qubit in Secs. \ref{sec:pure} and \ref{sec:mixed}. 
The optimal measurements are classified into three types by a value of error margin. 
For an arbitrary error margin, complete analytical results can be obtained. 
%%%%%%%%%%%%%%%%%%%%%%%%%%%%%%%%%%%%%%%%%%%%%%%%%%%%%%%%%%%%%%%%%%%%%%%%%%%%%%
%%% Formulation %%%
%%%%%%%%%%%%%%%%%%%%%%%%%%%%%%%%%%%%%%%%%%%%%%%%%%%%%%%%%%%%%%%%%%%%%%%%%%%%%%
\section{Formulation of problem} \label{sec:formulation}

We consider the discrimination problem among three quantum states, $\rho_{1}$, $\rho_{2}$, and $\rho_{3}$, with occurrence probabilities $\eta_{1}$, $\eta_{2}$, and $\eta_{3}$, respectively. 
Measurement is described by a positive operator-valued measure (POVM), which consists of four elements $\{ E_{0}, E_{1}, E_{2}, E_{3} \}$. A measurement outcome labeled $i$ = 1, 2, or 3 means that the given state is identified with state $\rho_{i}$, and the element $E_{0}$ produces the inconclusive result. 

The joint probability $P_{\rho_{i}, E_{\mu}}$ that the given state is $\rho_{i}$ 
and the measurement outcome is $\mu$ is given by
\begin{align*}
P_{\rho_{i}, E_{\mu}} = \eta_{i}\tr{E_{\mu}\rho_{i}}.
\end{align*}
The discrimination success probability $p_{\circ}$ and the mean probability of error $p_{\times}$ are given by
\begin{align}
p_{\circ} & \equiv P_{\rho_{1}, E_{1}} + P_{\rho_{2}, E_{2}} + P_{\rho_{3}, E_{3}} \nonumber \\
& = \eta_{1}\tr{E_{1}\rho_{1}} + \eta_{2}\tr{E_{2}\rho_{2}} + \eta_{3}\tr{E_{3}\rho_{3}}, \\
\nonumber \\
p_{\times} & \equiv P_{\rho_{2}, E_{1}} + P_{\rho_{3}, E_{1}} + P_{\rho_{1}, E_{2}} + P_{\rho_{3}, E_{2}} + P_{\rho_{1}, E_{3}} + P_{\rho_{2}, E_{3}} \nonumber \\
& = \tr{E_{1}\left(\eta_{2}\rho_{2} + \eta_{3}\rho_{3}\right)} + \tr{E_{2}\left(\eta_{1}\rho_{1} + \eta_{3}\rho_{3}\right)} \nonumber \\
& \ \ \ \ + \tr{E_{3}\left(\eta_{1}\rho_{1} + \eta_{2}\rho_{2}\right)}.
\end{align}

Our task is to maximize the discrimination success probability $p_{\circ}$ under the conditions
\begin{subequations}
\label{eq:Primal_problem}
\begin{align}
& E_{0} \geq 0, \ E_{1} \geq 0, \ E_{2} \geq 0, \ E_{3} \geq 0, \label{eq:POVM_positivity} \\
\nonumber \\
& E_{0} + E_{1} + E_{2} + E_{3} = \mbold{1}, \label{eq:POVM_completeness_relation} \\
\nonumber \\
& p_{\times} \leq m, \label{eq:error_margin_condition}
\end{align}
\end{subequations}
where Eqs. (\ref{eq:POVM_positivity}) and (\ref{eq:POVM_completeness_relation}) are the usual conditions for a POVM, and Eq. (\ref{eq:error_margin_condition}) is the condition that the mean probability of error $p_{\times}$ should not exceed a certain error margin $m$ ($0 \leq m \leq 1$). 
It is clear that unambiguous discrimination is formulated as the case of $m = 0$, while minimum-error discrimination corresponds to the case of $m = 1$. 
Thus, this scheme continuously interpolates two standard settings of quantum state discrimination. 

It is easy to see that this optimization problem can be formulated as one of semidefinite programming (SDP)\cite{Vandenberghe96}. 
For applications of SDP to quantum-state discrimination, see Refs. \cite{Eldar_Megretski_Verghese03} and \cite{Eldar_IEEE03}. 
 
Suppose a Hermitian operator $Y$ and a real number $y$ exist such that 
\begin{subequations}
\label{eq:Dual_problem}
\begin{align}
& Y \geq 0, \label{eq:Y_positivity}\\
\nonumber \\
& Y \geq \eta_{1}\rho_{1} - y\left(\eta_{2}\rho_{2} + \eta_{3}\rho_{3}\right), \label{eq:Y1_positivity} \\
\nonumber \\
& Y \geq \eta_{2}\rho_{2} - y\left(\eta_{1}\rho_{1} + \eta_{3}\rho_{3}\right), \label{eq:Y2_positivity} \\
\nonumber \\
& Y \geq \eta_{3}\rho_{3} - y\left(\eta_{1}\rho_{1} + \eta_{2}\rho_{2}\right), \label{eq:Y3_positivity} \\
\nonumber \\
& y \geq 0. \label{eq:real_number_y_positivity}
\end{align}
\end{subequations}

It is easy to show that
\begin{align}
d \equiv \tr{Y} + ym,
\end{align}
gives an upper bound for the discrimination success probability $p_{\circ}$, because
\begin{align*}
p_{\circ} & = \tr{E_{1}\eta_{1}\rho_{1}} + \tr{E_{2}\eta_{2}\rho_{2}} + \tr{E_{3}\eta_{3}\rho_{3}} \\
& \leq \tr{E_{1}\left[Y + y\left(\eta_{2}\rho_{2} + \eta_{3}\rho_{3}\right)\right]} \\
& \ \ \ \ + \tr{E_{2}\left[Y + y\left(\eta_{1}\rho_{1} + \eta_{3}\rho_{3}\right)\right]} \\
& \ \ \ \ + \tr{E_{3}\left[Y + y\left(\eta_{1}\rho_{1} + \eta_{2}\rho_{2}\right)\right]} \\
& = \tr{\left(E_{1} + E_{2} + E_{3}\right)Y} + yp_{\times} \\
& \leq \tr{Y} + ym = d.
\end{align*}
It is clear that this upper bound is attained if and only if the following relations hold:
\begin{subequations}
\begin{align}
& E_{1}\left[Y - \eta_{1}\rho_{1} + y\left(\eta_{2}\rho_{2} + \eta_{3}\rho_{3}\right)\right] = 0, \label{eq:attainability_condition1} \\ 
\nonumber \\
& E_{2}\left[Y - \eta_{2}\rho_{2} + y\left(\eta_{1}\rho_{1} + \eta_{3}\rho_{3}\right)\right] = 0, \label{eq:attainability_condition2} \\
\nonumber \\
& E_{3}\left[Y - \eta_{3}\rho_{3} + y\left(\eta_{1}\rho_{1} + \eta_{2}\rho_{2}\right)\right] = 0, \label{eq:attainability_condition3} \\
\nonumber \\
& E_{0}Y = 0, \label{eq:attainability_condition4} \\
\nonumber \\
& y\left(m - p_{\times}\right) = 0. \label{eq:attainability_condition5}
\end{align}
\end{subequations}
These five relations are called attainability conditions hereafter. 

Thus, optimal solutions can be obtained if we find a POVM $\{E_\mu\}$, an operator $Y$, 
and a real number $y$ which satisfy conditions Eqs. (\ref{eq:POVM_positivity})-(\ref{eq:error_margin_condition}), Eqs. (\ref{eq:Y_positivity})-(\ref{eq:real_number_y_positivity}), 
and Eqs. (\ref{eq:attainability_condition1})-(\ref{eq:attainability_condition5}).
Minimizing $d$ under conditions Eqs. (\ref{eq:Y_positivity})-(\ref{eq:real_number_y_positivity}) is called the dual problem, whereas the original problem of maximizing $p_{\circ}$ under conditions Eqs. (\ref{eq:POVM_positivity})-(\ref{eq:error_margin_condition}) is referred to as the primal problem.

Our strategy to obtain optimal solutions is as follows: 
We start with the dual problem. Minimization of $d$ is sometimes performed by adding 
extra conditions other than Eqs. (\ref{eq:Y_positivity})-(\ref{eq:real_number_y_positivity}) 
on $Y$ and $y$. We then construct POVM $\{E_\mu\}$ so that Eqs. (\ref{eq:POVM_positivity})-(\ref{eq:error_margin_condition})
and Eqs. (\ref{eq:attainability_condition1})-(\ref{eq:attainability_condition5}) are fulfilled. 
In this way, we obtain the maximum discrimination success probability and the optimal measurement that attains it. Note that the extra conditions in the dual problem do not hamper the strictness 
of our optimization. 

%%%%%%%%%%%%%%%%%%%%%%%%%%%%%%%%%%%%%%%%%%%%%%%%%%%%%%%%%%%%%%%%%%%%%%%%%%%%%%
%%% Pure state %%%
%%%%%%%%%%%%%%%%%%%%%%%%%%%%%%%%%%%%%%%%%%%%%%%%%%%%%%%%%%%%%%%%%%%%%%%%%%%%%%
\section{Three symmetric pure states of a qubit} \label{sec:pure}

In this section, we consider the case of three symmetric pure states of a qubit $\rho_{i} = \ket{\phi_{i}}\bra{\phi_{i}}$, where we assume that 
the occurrence probabilities $\eta_{i}$ are equal and the absolute values of all mutual inner products are the same, $|\braket{\phi_1}{\phi_2}|=|\braket{\phi_2}{\phi_3}|=|\braket{\phi_3}{\phi_1}|$.
It is convenient to use the Bloch vector representation for $\rho_{i}$ and other operators.
The three density operators are given by 
\begin{align*}
\rho_{i} = \frac{1 + \mbold{n}_{i}\cdot\mbold{\sigma}}{2} \ \ \ (i = 1, 2, 3),
\end{align*}
where $\mbold{\sigma} = (\sigma_{x}, \sigma_{y}, \sigma_{z})$ are Pauli's matrices.
The mutual inner products between the Bloch vectors $\mbold{n}_{i}$ satisfy
\begin{align}
\mbold{n}_{i}\cdot\mbold{n}_{j} \equiv \left\{
\begin{array}{ll}
1 & (i = j), \\
\gamma & (i \neq j),
\end{array}
\right.
\end{align}
where $\gamma$ is the only one parameter that characterizes the three symmetric pure states 
of a qubit, and it is in the range of $-1/2 \leq \gamma \leq 1$.

Since $Y$ is a 2 $\times$ 2 Hermitian operator, it can be written as 
\begin{align*}
Y = \frac{\alpha + \mbold{\beta}\cdot\mbold{\sigma}}{2}.
\end{align*}

The conditions given by Eqs. (\ref{eq:Y_positivity})-(\ref{eq:Y3_positivity}) imply that operators
\begin{align*}
Y_{1} & \equiv Y - \frac{1}{3}\left[\rho_{1} - y\left(\rho_{2} + \rho_{3}\right)\right], \\
\\
Y_{2} & \equiv Y - \frac{1}{3}\left[\rho_{2} - y\left(\rho_{1} + \rho_{3}\right)\right], \\
\\
Y_{3} & \equiv Y - \frac{1}{3}\left[\rho_{3} - y\left(\rho_{1} + \rho_{2}\right)\right],
\end{align*}
and $Y$ are positive semidefinite. By using the Bloch vector representation, operators $Y_{1}$, $Y_{2}$, and $Y_{3}$ are expressed as 
\begin{align*}
Y_{1} & = \frac{1}{2}\left(\alpha - \frac{1 - 2y}{3} + \left(\mbold{\beta} - \mbold{a}_{1}\right)\cdot\mbold{\sigma}\right), \\
\\ 
Y_{2} & = \frac{1}{2}\left(\alpha - \frac{1 - 2y}{3} + \left(\mbold{\beta} - \mbold{a}_{2}\right)\cdot\mbold{\sigma}\right), \\
\\
Y_{3} & = \frac{1}{2}\left(\alpha - \frac{1 - 2y}{3} + \left(\mbold{\beta} - \mbold{a}_{3}\right)\cdot\mbold{\sigma}\right), 
\end{align*}
where we introduced three vectors $\mbold{a}_{1}$, $\mbold{a}_{2}$, and $\mbold{a}_{3}$ defined 
by 
\begin{align}
\mbold{a}_{1} & \equiv \frac{1}{3}\left[\mbold{n}_{1} - y\left(\mbold{n}_{2} + \mbold{n}_{3}\right)\right], \label{eq:vector_a1} \\
\nonumber \\
\mbold{a}_{2} & \equiv \frac{1}{3}\left[\mbold{n}_{2} - y\left(\mbold{n}_{1} + \mbold{n}_{3}\right)\right], \label{eq:vector_a2} \\
\nonumber \\
\mbold{a}_{3} & \equiv \frac{1}{3}\left[\mbold{n}_{3} - y\left(\mbold{n}_{1} + \mbold{n}_{2}\right)\right]. \label{eq:vector_a3}
\end{align}
Since the smaller eigenvalue of operators $Y_{1}$, $Y_{2}$, $Y_{3}$, and $Y$ are positive, we obtain the following four inequalities for $\alpha$ and $\mbold{\beta}$:
\begin{align*}
& \alpha \geq \frac{1 - 2y}{3} + \left|\mbold{\beta} - \mbold{a}_{1}\right|, \\
\nonumber \\
& \alpha \geq \frac{1 - 2y}{3} + \left|\mbold{\beta} - \mbold{a}_{2}\right|, \\
\nonumber \\
& \alpha \geq \frac{1 - 2y}{3} + \left|\mbold{\beta} - \mbold{a}_{3}\right|, \\
\nonumber \\
& \alpha \geq \left|\mbold{\beta}\right|.
\end{align*}
Moreover, the upper bound $d$ can be rewritten as
\begin{align*}
d = \tr{\frac{\alpha + \mbold{\beta}\cdot\mbold{\sigma}}{2}} + ym = \alpha + ym.
\end{align*}

In terms of parameters $\{ y, \alpha, \mbold{\beta} \}$, the dual problem takes the following form: minimize
\begin{subequations}
\label{eq:formulation_of_alpha_and_beta}
\begin{align}
d = \alpha + ym,
\end{align}
subject to
\begin{align}
& y \geq 0, \\
\nonumber \\
& \alpha \geq \frac{1 - 2y}{3} + \left|\mbold{\beta} - \mbold{a}_{1}\right|, \label{eq:Y1_eigenvalue_positivity} \\
\nonumber \\
& \alpha \geq \frac{1 - 2y}{3} + \left|\mbold{\beta} - \mbold{a}_{2}\right|, \label{eq:Y2_eigenvalue_positivity} \\
\nonumber \\
& \alpha \geq \frac{1 - 2y}{3} + \left|\mbold{\beta} - \mbold{a}_{3}\right|, \label{eq:Y3_eigenvalue_positivity} \\
\nonumber \\
& \alpha \geq \left|\mbold{\beta}\right|. \label{eq:Y_eigenvalue_positivity}
\end{align}
\end{subequations}

In what follows, we present the main results first, leaving their derivation to subsequent subsections. The parameter space is divided into the following three domains,
\begin{align*}
\begin{array}{ll}
m_{c} \leq m \leq 1 & (\mbox{minimum-error domain}), \\
m'_{c} \leq m \leq m_{c} & (\mbox{intermediate domain}), \\
0 \leq m \leq m'_{c} & (\mbox{linear domain}),
\end{array}
\end{align*}
where two critical error margins $m_{c}$ and $m'_{c}$ are defined by
\begin{align}
m_{c} & \equiv \frac{1}{3}\left(2 - \sqrt{\frac{2\left(1 - \gamma\right)}{3}}\right), \label{eq:mc_pure} \\
\nonumber \\
m'_{c} & \equiv \frac{1}{3}\left(1 - \sqrt{\frac{1 + 2\gamma}{3}}\right). \label{eq:mc'_pure}
\end{align}

The maximum discrimination success probability in each domain is found to be
\begin{align}
p_{\max} = \left\{
\begin{array}{ll}
\frac{1}{3}\left(1 + \sqrt{\frac{2\left(1 - \gamma\right)}{3}}\right) & (m_{c} \leq m \leq 1), \\
\frac{1}{2}\left(m + A + \sqrt{3A\left(2m - A\right)}\right) & (m'_{c} \leq m \leq m_{c}), \\
2m & (0 \leq m \leq m'_{c}), 
\end{array}
\right.
\end{align}
where $A$ is a positive constant defined by
\begin{align}
A = \frac{1}{2}\left(1 - \sqrt{\frac{1 + 2\gamma}{3}}\right). \label{eq:A}
\end{align}

%%%%%%%%%%%%%%%%%%%%%%%%%%%%%%%%%%%%%%%%%%%%%%%%%%%%%%%%%%%%%%
%%% Minimum error domain %%%
%%%%%%%%%%%%%%%%%%%%%%%%%%%%%%%%%%%%%%%%%%%%%%%%%%%%%%%%%%%%%%
\subsection{Minimum-error domain}

Suppose the error margin is so large that the constraint on the probability of error is inactive, which means that  the probability of error $p_{\times}$ in the optimal measurement is 
strictly smaller than the error margin $m$. Then, the optimal measurement is that of minimum-error discrimination. Hereafter, the domain where this is the case is called minimum-error domain.

In the minimum-error domain, the condition given by Eq. (\ref{eq:attainability_condition5}) implies that parameter $y$ should be 0 since $p_{\times} < m$. 
From this consideration, we can rewrite the discrimination problem as follows: minimize
\begin{subequations}
\begin{align}
d = \alpha,
\end{align}
subject to
\begin{align}
\alpha & \geq \frac{1}{3} + \left|\mbold{\beta} - \frac{1}{3}\mbold{n}_{1}\right|, \label{eq:Y1_eigenvalue_positivity_ME} \\
\nonumber \\
\alpha & \geq \frac{1}{3} + \left|\mbold{\beta} - \frac{1}{3}\mbold{n}_{2}\right|, \label{eq:Y2_eigenvalue_positivity_ME} \\
\nonumber \\
\alpha & \geq \frac{1}{3} + \left|\mbold{\beta} - \frac{1}{3}\mbold{n}_{3}\right|. \label{eq:Y3_eigenvalue_positivity_ME}
\end{align}
\end{subequations}

The condition given by Eq. (\ref{eq:Y_eigenvalue_positivity}) can be omitted, since 
Eq. (\ref{eq:Y_eigenvalue_positivity}) is a consequence of the conditions Eqs. (\ref{eq:Y1_eigenvalue_positivity_ME})-(\ref{eq:Y3_eigenvalue_positivity_ME}). This is obvious from the inequalities
\begin{align*}
\frac{1}{3} + \left|\mbold{\beta} - \frac{1}{3}\mbold{n}_{i}\right| \geq \left|\mbold{\beta}\right| \ \ \ (i = 1, 2, 3),
\end{align*}
which are obtained by applying the triangle inequality on the right-hand side of Eqs. (\ref{eq:Y1_eigenvalue_positivity_ME})-(\ref{eq:Y3_eigenvalue_positivity_ME}).

We construct a solution where all POVM elements $E_{1}$, $E_{2}$, and $E_{3}$ are of rank 1. Thus, we assume that equality holds in Eqs. (\ref{eq:Y1_eigenvalue_positivity_ME})-(\ref{eq:Y3_eigenvalue_positivity_ME}) as follows:
\begin{subequations}
\begin{align}
\alpha & = \frac{1}{3} + \left|\mbold{\beta} - \frac{1}{3}\mbold{n}_{1}\right|, \label{eq:alpha_beta_equation1_ME} \\
\nonumber \\
\alpha & = \frac{1}{3} + \left|\mbold{\beta} - \frac{1}{3}\mbold{n}_{2}\right|, \label{eq:alpha_beta_equation2_ME} \\
\nonumber \\
\alpha & = \frac{1}{3} + \left|\mbold{\beta} - \frac{1}{3}\mbold{n}_{3}\right|. \label{eq:alpha_beta_equation3_ME}
\end{align}
\end{subequations}
By solving Eqs. (\ref{eq:alpha_beta_equation1_ME})-(\ref{eq:alpha_beta_equation3_ME}), it turns out that vector $\mbold{\beta}$ is given by
\begin{align}
\mbold{\beta} = \frac{1}{9}\left(\mbold{n}_{1} + \mbold{n}_{2} + \mbold{n}_{3}\right), \label{eq:beta_pure_ME}
\end{align}
and $\alpha$ is given by
\begin{align}
\alpha = d = \frac{1}{3}\left(1 + \sqrt{\frac{2\left(1 - \gamma\right)}{3}}\right). \label{eq:upper_bound_pure_ME}
\end{align}
We construct the set of POVM $\{ E_\mu \}$ which attains the upper bound $d$ given by Eq. (\ref{eq:upper_bound_pure_ME}). 
The element $E_0$ for the inconclusive result is taken to be 0 in this domain and 
the attainability condition Eq. (\ref{eq:attainability_condition4}) is satisfied. 
The attainability conditions given by Eqs. (\ref{eq:attainability_condition1})-(\ref{eq:attainability_condition3}) require that POVM elements $E_{1}$, $E_{2}$, and $E_{3}$ take the following form:
\begin{align*}
E_{i} = C_{i}\frac{1 - \mbold{e}_{i}\cdot\mbold{\sigma}}{2} \ \ \ (i = 1, 2, 3).
\end{align*}
Here, we defined three unit vectors, 
\begin{align*}
\mbold{e}_{i} \equiv \frac{\mbold{\beta} - \frac{1}{3}\mbold{n}_{i}}{\left|\mbold{\beta} - \frac{1}{3}\mbold{n}_{i}\right|} \ \ \ (i = 1, 2, 3),
\end{align*}
where vector $\mbold{\beta}$ is given in Eq. (\ref{eq:beta_pure_ME}).

We determine the coefficients $C_{i}$ $(i = 1, 2, 3)$ so that the conditions given by Eqs. (\ref{eq:POVM_positivity})-(\ref{eq:error_margin_condition}) are satisfied. 
From the completeness relation of POVM given by Eq. (\ref{eq:POVM_completeness_relation}), we obtain the two relations for the coefficients $C_{i}$ as follows: 
\begin{align}
\frac{1}{2}\left(C_{1} + C_{2} + C_{3}\right) & = 1, \label{eq:part_of_coefficient_ME} \\
\nonumber \\
C_{1}\mbold{e}_{1} + C_{2}\mbold{e}_{2} + C_{3}\mbold{e}_{3} & = 0. \label{eq:part_of_vector_ME}
\end{align}
In Eq. (\ref{eq:part_of_vector_ME}), each of the three vectors $\mbold{e}_{1}$, $\mbold{e}_{2}$, and $\mbold{e}_{3}$ is expressed by the three Bloch vectors $\mbold{n}_{1}$, $\mbold{n}_{2}$, and $\mbold{n}_{3}$. From the linear independence of the three Bloch vectors $\mbold{n}_{1}$, $\mbold{n}_{2}$, and $\mbold{n}_{3}$, we find $C_{1} = C_{2} = C_{3}$. Therefore, the coefficient $C_{i}$ $(i = 1, 2, 3)$ is given by
\begin{align*}
C_{1} = C_{2} = C_{3} = \frac{2}{3}.
\end{align*}
It is clear that the set of POVM $\{ E_{1}, E_{2}, E_{3} \}$ satisfies Eq. (\ref{eq:POVM_positivity}) since $C_{i} > 0$.

By the remaining condition Eq. (\ref{eq:error_margin_condition}), we find 
\begin{align}
p_{\times} = \frac{1}{3}\left(2 - \sqrt{\frac{2\left(1 - \gamma\right)}{3}}\right) \leq m,
\end{align}
where the mean probability of error $p_{\times}$ is calculated by using the POVM constructed above.

Thus, if the error margin $m$ is in the range $m_{c} \leq m \leq 1$, the upper bound of Eq. (\ref{eq:upper_bound_pure_ME}) is attained and the maximum discrimination success probability is given by
\begin{align}
p_{\max} = \frac{1}{3}\left(1 + \sqrt{\frac{2\left(1 - \gamma\right)}{3}}\right).
\end{align}

%%%%%%%%%%%%%%%%%%%%%%%%%%%%%%%%%%%%%%%%%%%%%%%%%%%%%%%%%%%%%%
%%% Linear and Intermediate domain %%%
%%%%%%%%%%%%%%%%%%%%%%%%%%%%%%%%%%%%%%%%%%%%%%%%%%%%%%%%%%%%%%
\subsection{Linear and intermediate domain}

In this subsection, we construct a solution where all POVM elements $E_{1}$, $E_{2}$, $E_{3}$, and $E_{0}$ are nonzero. 
The attainability conditions given by Eqs. (\ref{eq:attainability_condition1})-(\ref{eq:attainability_condition4}) imply that operators $Y_{1}$, $Y_{2}$, $Y_{3}$, and $Y$ are rank 1 at most. 
That is, the smaller eigenvalues of operators $Y_{1}$, $Y_{2}$, $Y_{3}$, and $Y$are all 0. 
Thus, we assume that equality holds in Eqs. (\ref{eq:Y1_eigenvalue_positivity})-(\ref{eq:Y_eigenvalue_positivity}) as follows:
\begin{subequations}
\begin{align}
\alpha & = \frac{1 - 2y}{3} + \left|\mbold{\beta} - \mbold{a}_{1}\right|, \label{eq:alpha_beta_equation1} \\
\nonumber \\
\alpha & = \frac{1 - 2y}{3} + \left|\mbold{\beta} - \mbold{a}_{2}\right|, \label{eq:alpha_beta_equation2} \\
\nonumber \\
\alpha & = \frac{1 - 2y}{3} + \left|\mbold{\beta} - \mbold{a}_{3}\right|, \label{eq:alpha_beta_equation3} \\
\nonumber \\
\alpha & = \left|\mbold{\beta}\right|. \label{eq:alpha_beta_equation4}
\end{align}
\end{subequations}

Solving Eqs. (\ref{eq:alpha_beta_equation1})-(\ref{eq:alpha_beta_equation4}) requires a rather complicated and long calculation. It turns out that parameter $y$ must satisfy $1/2 \leq y \leq 2$, vector $\mbold{\beta}$ is given by
\begin{align}
\mbold{\beta} = \frac{1}{2}\left(1 \pm \sqrt{\frac{3}{1 + 2\gamma}}\right)\frac{\left(y - 2\right)y}{1 - 2y}\left(\frac{\mbold{n}_{1} + \mbold{n}_{2} + \mbold{n}_{3}}{3}\right), \label{eq:vector_beta_pure}
\end{align}
and $\alpha$ is given by
\begin{align}
\alpha = \left|\mbold{\beta}\right| = \frac{1}{2}\left(1 \pm \sqrt{\frac{1 + 2\gamma}{3}}\right)\frac{\left(y - 2\right)y}{1 - 2y}. \label{eq:alpha_pure}
\end{align}
Therefore, we obtain an upper bound for the discrimination success probability as a function of parameter $y$:
\begin{align}
d = \alpha + ym = \frac{1}{2}\left(1 \pm \sqrt{\frac{1 + 2\gamma}{3}}\right)\frac{\left(y - 2\right)y}{1 - 2y} + my. \label{eq:upper_bound_d_pure}
\end{align}
As for the double signs in the above equation, we take a negative one to obtain a smaller upper bound. 
Correspondingly, a negative sign is also taken in the double signs of Eqs. (\ref{eq:vector_beta_pure}) and (\ref{eq:alpha_pure}) hereafter. 

Now we determine parameter $y$ so that the upper bound $d$ is minimized. Let us begin by looking at how the upper bound $d$ changes in the range of parameter $1/2 \leq y \leq 2$. By differentiating the upper bound $d$ with respect to parameter $y$, we have
\begin{align*}
\frac{\partial}{\partial y}d = m - \frac{1}{2}A - \frac{3}{2}A\frac{1}{\left(2y - 1\right)^{2}}.
\end{align*} 
A positive constant $A$ is defined by Eq. (\ref{eq:A}).
If the error margin is $m < A/2$, it turns out that the upper bound $d$ is a monotone decreasing function with respect to parameter $y$. 
In this case, the upper bound $d$ takes the minimum $d_{\min} = 2m$ at $y = 2$. 
On the other hand, in the case of $A/2 \leq m$, the quadratic equation for 
a parameter $y$ produced by $\frac{\partial}{\partial y}d = 0$ has the roots $y_{e}$ given by
\begin{align}
y_{e} = \frac{1}{2}\left(1 \pm \sqrt{\frac{3A}{2m - A}}\right). \label{eq:ye_pure}
\end{align}
As for the double signs of the roots $y_{e}$, we take a positive one since $1/2 \leq y \leq 2$.

Here, we consider the two cases. 
One is the case of $y_{e} \geq 2$, which can be rewritten as a range of error margin $m$, 
\begin{align}
m \leq \frac{2}{3}A \equiv m'_{c}, \label{eq:error_margin_range_pure_LD}
\end{align}
where $m'_{c}$ on the right-hand side is defined in Eq. (\ref{eq:mc'_pure}). 
This domain is called the linear domain. 
In linear domain, the upper bound $d$ is minimum at $y = 2$. 
Moreover, $Y = 0$ since $\mbold{\beta}$ and $\alpha$ become 0. Therefore, the upper bound $d$ in linear domain is given by
\begin{align}
d = 2m, \label{eq:upper_bound_pure_LD}
\end{align}
which is linear with respect to the error margin $m$. 

The other is the case of $y_{e} < 2$, where we have 
\begin{align}
m'_{c} \equiv \frac{2}{3}A < m. \label{eq:left_side_of_range_pure_ID}
\end{align}
This domain is called the intermediate domain. 
In the intermediate domain, the upper bound $d$ is minimum at $y = y_{e}$. 
Using $Y$ and $y$, which minimize an upper bound for the discrimination success probability, we obtain the minimum of the upper bound $d = \tr{Y} + ym$ to be 
\begin{align}
d = \frac{1}{2}\left(m + A + \sqrt{3A\left(2m - A\right)}\right). \label{eq:upper_bound_pure_ID}
\end{align}

In the following subsections, we construct the optimal POVM to achieve the obtained upper bound of each domain.
%%%%%%%%%%%%%%%%%%%%%%%%%%%%%%%%%%%%%%%%%%%%%%%%%%
%%% Linear domain %%%
%%%%%%%%%%%%%%%%%%%%%%%%%%%%%%%%%%%%%%%%%%%%%%%%%%
\subsubsection{Linear domain} 

In linear domain, we have $Y = 0$ since $\mbold{\beta} = \alpha = 0$. 
This shows that the attainability condition given by Eq. (\ref{eq:attainability_condition4}) does not give any restriction on POVM element $E_{0}$. 
The attainability conditions given by Eqs. (\ref{eq:attainability_condition1})-(\ref{eq:attainability_condition3}) require that POVM elements $E_{1}$, $E_{2}$, and $E_{3}$ take the following form:
\begin{align*}
E_{i} = C \frac{1 + \mbold{e}_{i}\cdot\mbold{\sigma}}{2} \ \ \ (i = 1, 2, 3).
\end{align*}
Here, we assumed that the coefficient $C$ does not depend on index $i$ $(i = 1, 2, 3)$ and the vectors $\mbold{e}_{i}$ are defined by 
\begin{align*}
\mbold{e}_{i} & \equiv \mbold{a}_{i} \ \ \ (i = 1, 2, 3),
\end{align*}
where $\mbold{a}_{i}$ are given in Eqs. (\ref{eq:vector_a1})-(\ref{eq:vector_a3}) with $y = 2$. 

The question is whether the coefficient $C$ can be chosen so that the set $\{ E_{1}, E_{2}, E_{3} \}$ satisfies the conditions given by Eqs. (\ref{eq:POVM_positivity})-(\ref{eq:error_margin_condition}), and (\ref{eq:attainability_condition5}). 
The conditions given in Eqs. (\ref{eq:error_margin_condition}) and (\ref{eq:attainability_condition5}) are reduced to $p_{\times} = m$ since $y = 2$. 
Calculating $p_{\times}$ by the POVM constructed above, we find that the coefficient $C$ is given by 
\begin{align*}
C = \frac{3m}{1 - \gamma},
\end{align*}
and $C$ is positive since $-1/2 \leq \gamma \leq 1$. 
This shows that POVM elements $E_{1}$, $E_{2}$, and $E_{3}$ satisfy Eq. (\ref{eq:POVM_positivity}). 
In addition, POVM element $E_{0}$ is obtained from the completeness relation of POVM given in Eq. (\ref{eq:POVM_completeness_relation}) as follows:
\begin{align}
E_{0} & = \mbold{1} - \left(E_{1} + E_{2} + E_{3}\right) \nonumber \\
& = 1 - \frac{9m}{2\left(1 - \gamma\right)} + \frac{3m}{2\left(1 - \gamma\right)}\left(\mbold{n}_{1} + \mbold{n}_{2} + \mbold{n}_{3}\right)\cdot\mbold{\sigma}.
\end{align}
Here, the smaller eigenvalue of POVM element $E_{0}$, 
\begin{align*}
\lambda_{-} = 1 - \frac{9}{2\left(1 - \gamma\right)}\left(1 + \sqrt{\frac{1 + 2\gamma}{3}}\right)m,
\end{align*}
should be positive since $E_{0} \geq 0$. 
This is satisfied since the error margin $m$ is in the range $0 \leq m \leq m'_{c}$.

Therefore, if the error margin $m$ is in the range $0 \leq m \leq m'_{c}$, the upper bound of Eq. (\ref{eq:upper_bound_pure_LD}) is attained and the maximum discrimination success probability is given by
\begin{align}
p_{\max} = 2m.
\end{align}
%%%%%%%%%%%%%%%%%%%%%%%%%%%%%%%%%%%%%%%%%%%%%%%%%%
%%% Intermediate domain %%%
%%%%%%%%%%%%%%%%%%%%%%%%%%%%%%%%%%%%%%%%%%%%%%%%%%
\subsubsection{Intermediate domain}

We construct the set of POVM $\{ E_{0}, E_{1}, E_{2}, E_{3} \}$ which attains the upper bound $d$ given by Eq. (\ref{eq:upper_bound_pure_ID}). 
The attainability conditions given by Eq. (\ref{eq:attainability_condition1})-(\ref{eq:attainability_condition4}) require that POVM elements $E_{1}$, $E_{2}$, $E_{3}$, and $E_{0}$ take the form
\begin{align*}
E_{\mu} = C_{\mu}\frac{1 + \mbold{e}_{\mu}\cdot\mbold{\sigma}}{2} \ \ \ (\mu = 0, 1, 2, 3),
\end{align*}
where we defined
\begin{align*}
\mbold{e}_{0} \equiv \frac{\mbold{\beta}}{\left|\mbold{\beta}\right|}, \ \mbold{e}_{i} \equiv \frac{\mbold{\beta} - \mbold{a}_{i}}{\left|\mbold{\beta} - \mbold{a}_{i}\right|} \ \ \ (i = 1, 2, 3).
\end{align*}
Here, $\mbold{\beta}$ is given in Eq. (\ref{eq:vector_beta_pure}) with $y_{e}$ given by Eq. (\ref{eq:ye_pure}) substituted for $y$, and vectors $\mbold{a}_{i}$ $(i = 1, 2, 3)$ are given in Eqs. (\ref{eq:vector_a1}), (\ref{eq:vector_a2}), and (\ref{eq:vector_a3}) with $y = y_{e}$.

The completeness relation of POVM given by Eq. (\ref{eq:POVM_completeness_relation}) is now 
expressed as 
\begin{align}
\frac{1}{2}\left(C_{0} + C_{1} + C_{2} + C_{3}\right) & = 1, \label{eq:part_of_coefficient_ID} \\
\nonumber \\
C_{0}\mbold{e}_{0} + C_{1}\mbold{e}_{1} + C_{2}\mbold{e}_{2} + C_{3}\mbold{e}_{3} & = 0. \label{eq:part_of_vector_ID}
\end{align}
In Eq. (\ref{eq:part_of_vector_ID}), each of the four vectors $\mbold{e}_{1}$, $\mbold{e}_{2}$, $\mbold{e}_{3}$, and $\mbold{e}_{0}$ is expressed by the three Bloch vectors $\mbold{n}_{1}$, $\mbold{n}_{2}$, and $\mbold{n}_{3}$. From the linear independence of the three Bloch vectors $\mbold{n}_{1}$, $\mbold{n}_{2}$, and $\mbold{n}_{3}$, we can see that there is the relation $C_{1} = C_{2} = C_{3}$. Hereafter, we write $C$ for $C_{i}$. After a rather long calculation, we obtain
\begin{align*}
C & = \frac{A}{1 - \gamma}\left[2 + 3\left(m - A\right) + \sqrt{3A\left(2m - A\right)}\right], \\ 
\\
C_{0} & = 2 - 3C.
\end{align*}
The coefficients $C$ and $C_{0}$ should be positive to satisfy Eq. (\ref{eq:POVM_positivity}). 
This is satisfied since the error margin $m$ is in the range $m'_{c} \leq m \leq m_{c}$.

The remaining conditions are Eqs. (\ref{eq:error_margin_condition}) and (\ref{eq:attainability_condition5}), which are reduced to $p_{\times} = m$ since $1/2 \leq y \leq 2$. 
We can explicitly verify that the relation $p_{\times} = m$ holds after a long calculation by using the POVM constructed above. 
This is not a coincidence, but a consequence of how we determined parameter $y$. 
Parameter $y$ was determined so that the upper bound $d$ given by Eq. (\ref{eq:upper_bound_d_pure}) is minimized:
\begin{align*}
\frac{\partial}{\partial y}d = \frac{\partial}{\partial y}\tr{Y} + m = 0.
\end{align*}
Using the same argument given in Ref. \cite{Sugimoto09}, 
we can show that $\frac{\partial}{\partial y}\tr{Y} = -p_{\times}$, which means that minimization of $d$ leads to the relation $p_{\times} = m$. 

Thus, if the error margin $m$ is in the range $m'_{c} \leq m \leq m_{c}$, the upper bound of Eq. (\ref{eq:upper_bound_pure_ID}) is attained and the maximum discrimination success probability is given by
\begin{align}
p_{\max} = \frac{1}{2}\left(m + A + \sqrt{3A\left(2m - A\right)}\right).
\end{align}

%%%%%%%%%%%%%%%%%%%%%%%%%%%%%%%%%%%%%%%%%%%%%%%%%%%%%%%%%%%%%%%%%%%%%%%%%%%%%%
%%% Example %%%
%%%%%%%%%%%%%%%%%%%%%%%%%%%%%%%%%%%%%%%%%%%%%%%%%%%%%%%%%%%%%%%%%%%%%%%%%%%%%%
\subsection{Example}

We consider the set of states defined by the following Bloch vectors $\mbold{n}_{1}$, $\mbold{n}_{2}$, and $\mbold{n}_{3}$:
\begin{align*}
\mbold{n}_{1} = \left(
\begin{array}{c}
1 \\
0 \\
0 
\end{array}
\right), \ \mbold{n}_{2} = \left(
\begin{array}{c}
0 \\
1 \\
0
\end{array}
\right), \ \mbold{n}_{3} = \left(
\begin{array}{c}
0 \\
0 \\
1
\end{array}
\right).
\end{align*}
The mutual inner products between Bloch vectors is $\gamma = \mbold{n}_{i}\cdot\mbold{n}_{j} = 0$ $(i \neq j)$.

\begin{figure}
\includegraphics[width=7cm]{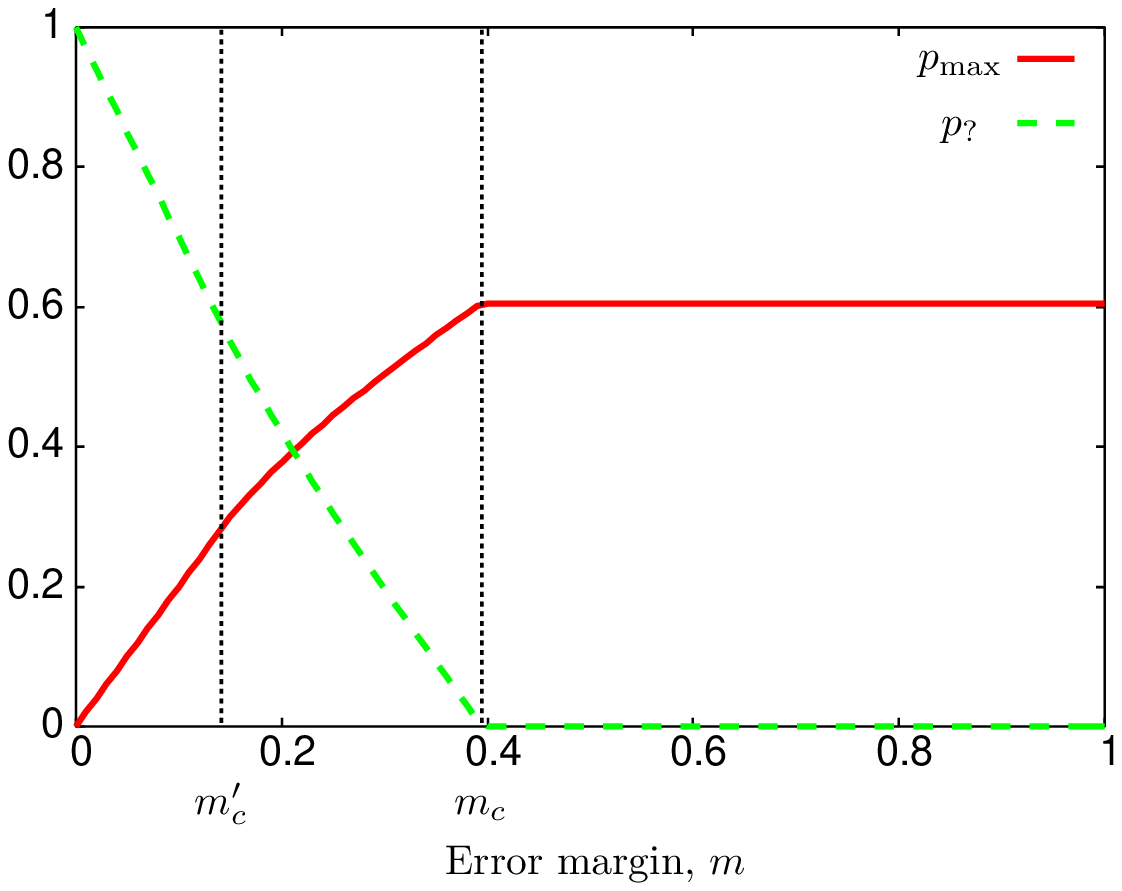}
\caption{(Color online) Maximum discrimination success probability $p_{\max}$  [solid (red) line] and probability of an inconclusive result $p_{?}$ [dashed (green) line] vs error margin $m$.}
\label{fig:example}
\end{figure}

Figure \ref{fig:example} displays the maximum discrimination success probability $p_{\max}$ and the probability of an inconclusive result $p_{?}$ as a function of the error margin $m$. 
As the error margin increases from 0 to 1, the type of optimal measurement varies in the following way: from the linear to the intermediate type at $m = m'_{c}$, and the minimum-error type at $m = m_{c}$. 
The maximum discrimination success probability $p_{\max}$ and the probability of an inconclusive result $p_{?}$ do not depend on the error margin $m$ in the range  $m_{c} \leq m \leq 1$.
Moreover, the curves of $p_{\max}$ and $p_{?}$ clearly show the border between the minimum-error and the intermediate domain, though the curves are smooth at $m = m'_{c}$.

In our previous paper \cite{Sugimoto09}, we analyzed discrimination with the error margin between two pure states with general occurrence probabilities and found that there is a domain where omitting one of the states to be discriminated is optimal if the error margin is sufficiently small. 
However, there is no such domain for the symmetric states considered in the present paper, where there is a symmetry with respect to interchange of the states.

%%%%%%%%%%%%%%%%%%%%%%%%%%%%%%%%%%%%%%%%%%%%%%%%%%%%%%%%%%%%%%%%%%%%%%%%%%%%%%
%%% Mixed state %%%
%%%%%%%%%%%%%%%%%%%%%%%%%%%%%%%%%%%%%%%%%%%%%%%%%%%%%%%%%%%%%%%%%%%%%%%%%%%%%%
\section{Three symmetric mixed states of a qubit} \label{sec:mixed}

In this section, we consider that three states to be discriminated, $\rho_{1}$, $\rho_{2}$, and $\rho_{3}$, are mixed and symmetric and satisfy the following conditions:
\begin{eqnarray*}
\tr{\rho^{2}_{1}} = \tr{\rho^{2}_{2}} = \tr{\rho^{2}_{3}} < 1, \ \tr{\rho_{1}\rho_{2}} = \tr{\rho_{2}\rho_{3}} = \tr{\rho_{3}\rho_{1}}.
\end{eqnarray*}
The mutual inner products between the Bloch vectors are parametrized as  
\begin{align}
\mbold{n}_{i}\cdot\mbold{n}_{j} \equiv \left\{
\begin{array}{ll}
r & (i = j), \\
\gamma & (i \neq j),
\end{array}
\right.
\end{align}
where $0 \leq r \leq 1$ and $-r/2 \leq \gamma \leq r$. 
The symmetric set of three mixed states of a qubit is characterized by two parameters, $r$ and $\gamma$.

The problem is formulated in exactly the same way as in Eq. (\ref{eq:formulation_of_alpha_and_beta}). 
The optimal solution can be obtained as in the case of pure states, 
though the calculation becomes much more complex. 

In what follows, we present the results of the case of three symmetric mixed states. To make the expressions simpler, we define 
\begin{align}
s & \equiv 1 - r, \\
t & \equiv 1 - \gamma. 
\end{align}
We obtain the maximum discrimination success probability in the case of three  symmetric mixed states of a qubit as 
\begin{align}
& p_{\max} = \nonumber \\
& \left\{
\begin{array}{ll}
\frac{1}{3}\left(1 + \sqrt{\frac{2\left(t - s\right)}{3}}\right) & (m_{c} \leq m \leq 1), \\
\\
\frac{1}{s + 2t}\Big(\frac{1}{2}m\left(s + 2t\right) + A\left(t - s\right) \\
+ \sqrt{3A\left(t - s\right)\left(m\left(s + 2t\right) - A\left(s + t\right)\right)}\Big) \\
& (m'_{c} \leq m \leq m_{c}), \\
\\
\frac{1}{s + t}\left(t + \sqrt{\frac{\left(t - s\right)\left(s + 2t\right)}{2}}\right)m & (0 \leq m \leq m'_{c}),
\end{array}
\right. \nonumber \\
\end{align}
where $A$ is given by
\begin{align}
A \equiv \frac{1}{2}\left(1 - \sqrt{\frac{r + 2\gamma}{3}}\right),
\end{align}
and $m_{c}$ and $m'_{c}$ are defined by
\begin{align}
m_{c} & \equiv \frac{1}{3}\left(2 - \sqrt{\frac{2\left(t - s\right)}{3}}\right), \\
\nonumber \\
m'_{c} & \equiv \frac{2}{3}\left(2 - \sqrt{\frac{2\left(t - s\right)}{s + 2t}}\right)A.
\end{align}
Note that, when $r = 1$, this reproduces the results obtained in Sec. \ref{sec:pure}.

%%%%%%%%%%%%%%%%%%%%%%%%%%%%%%%%%%%%%%%%%%%%%%%%%%%%%%%%%%%%%%%%%%%%%%%%%%%%%%
%%% Conclusion %%%
%%%%%%%%%%%%%%%%%%%%%%%%%%%%%%%%%%%%%%%%%%%%%%%%%%%%%%%%%%%%%%%%%%%%%%%%%%%%%%
\section{Concluding remarks} \label{sec:conclusion}

We have considered a state discrimination problem which interpolates minimum-error and unambiguous discriminations by introducing a margin for the probability of error. 
In the case of three symmetric pure states, we obtained the optimal success probability in a fully analytic from. 
We also showed that our method can be applied to the three symmetric mixed states. 
The optimal measurement is classified into three types. One of the three types of measurement is optimal, depending on the error margin. We note that this classification is also done by 
the rank of POVM element $E_{0}$: 
\begin{align*}
\mbox{rank}\left(E_{0}\right) = \left\{
\begin{array}{ll}
0 & (\mbox{minimum-error domain}) \\
1 & (\mbox{intermediate domain}) \\
2 & (\mbox{linear domain})
\end{array}
\right..
\end{align*}
When rank$\left(E_{0}\right) = 0$, the optimal measurement is that of minimum-error domain. However, physical implication of the difference between the case of rank$\left(E_{0}\right) = 1$ and the case of rank$\left(E_{0}\right) = 2$ is not very clear.

{\it Note added in proof.} Recently we became aware of two recent related works \cite{Herzogrecent,Baganrecent}. 
In both works, discrimination with a fixed rate of inconclusive results is applied to the set of symmetric states and it is argued that the results can be transformed to solutions for discrimination with error margin. 

%%%%%%%%%%%%%%%%%%%%%%%%%%%%%%%%%%%%%%%%%%%%%%%%%%%%%%%%%%%%%%%%%%%%%%%%%%%%%%
%%% Reference %%
%%%%%%%%%%%%%%%%%%%%%%%%%%%%%%%%%%%%%%%%%%%%%%%%%%%%%%%%%%%%%%%%%%%%%%%%%%%%%%

%%%%%%%%%%%%%%%%%%%%%%%%%%%%%%%%%%%%%%%%%%%%%%%%%%%%%%%%%%%%%%%%%%%%%%%%%%%%%%
\end{document}